\documentclass[aps,twocolumn,showpacs,groupedaddress]{revtex4}
\usepackage{graphicx}
\newcommand{\Tr}{\text{Tr}}

\begin{document}

\title{Excitation energy transfer in light-harvesting system: Effect of initial state}
\author{B. Cui$^1$, X. X. Yi$^{1,2}$,
 and C. H. Oh$^2$}
\affiliation{$^1$School of Physics and Optoelectronic Technology\\
Dalian University of Technology, Dalian 116024 China\\
$^2$Centre for Quantum Technologies and Department of Physics,
National University of Singapore, 117543, Singapore }

\date{\today}

\begin{abstract}
The light-harvesting is a problem of long interest.  It becomes
active again  in recent years stimulated  by suggestions of quantum
effects in energy transport. Recent experiments  found evidence that
BChla 1 and BChla 6  are the first to be excited in the
Fenna-Matthews-Olson(FMO) protein, theoretical studies, however, are
mostly restricted  to consider the exciton in BChla 1 initially. In
this paper, we study the energy transport in the FMO complex by
taking different  initial states into account. Optimizations are
performed for the decoherence rates as to maximal transport
efficiency. Dependence of the energy transfer efficiency on the
initial states is given and discussed. Effects of fluctuations in
the site energies and couplings are also examined.

\end{abstract}

 \maketitle
\section{introduction}

The knowledge of  photosynthetic processes in bacteria and plants
may provide us deep insights into designing efficiently and robustly
artificial light harvesting systems, where the initial stages in the
conversion of solar energy into chemical and other forms of energy
can be described by exciton dynamics with trapping and decoherence.
Recent experimental and computational studies suggest that
environmental noise can assist the exciton transport and can be
optimized for maximal energy transfer
efficiency\cite{engel07,lee07,panit10,collini10,
collini09,gaab04,olaya08,fassioli09,plenio08,rebentrost09,caruso09,cao09,yang10}.
In particular, special attention  has been focused on the so-called
Fenna-Matthews-Olson (FMO) complex that promotes energy transfer
from the main light-harvesting complex towards the chemical reaction
center in green sulfur bacteria.

As a small protein in green sulfur bacterium, the FMO complex is a
highly efficient energy transfer wire that connects
chlorosomes(i.e., light collecting pigment arrays) to photosynthetic
reaction centers. In structure, the FMO protein is a trimer, each
monomer contains seven bacteriochlorophyll-a (BChla) molecules
embedded within a protein scaffold and is believed to function
independently\cite{wen09}. We shall refer to the individual BChla
molecules as sites in this paper. Recent studies have determined the
orientation of the FMO complex within the inter-membrane region
between the chlorosome antenna and reaction
center\cite{wen09,adolphs06},  evidence that photosynthetic systems
 exhibit quantum coherence at ambient temperature has been found, but
the significance of these discoveries for biological energy transfer
remains unclear. It is quite certain that the reaction center is
strongly coupled to BChla 3 and that the excitation energy enters an
FMO monomer from the chlorosomes via BChla 1 or BChla 6.

Although a preliminary understanding of the complex dynamics of
excitation transfer has been made, the theory  is still in its
infancy. For example, most studies are restricted to one exciton
limit and the initial exciton is supposed in only one of the seven
sites. This paper addresses two open questions concerning the
relevance of excitonic coherence to photosynthetic energy transfer.
First, we present a theoretical framework which elucidates how
decoherence could assist the excitation transfer, we optimize the
decoherence rates for maximal transfer efficiency and study the
robustness of excitation transfer against the fluctuation of the
on-site  energy. Second, we present a study on the coherence in the
initially excited states by showing the dependence of the transfer
efficiency on the initial states.

This paper is organized as follows. In Sec. {\rm II}, we introduce
the model  to describe the FMO complex and   the Markovian master
equation  for the dynamics of the exciton transport. In Sec. {\rm
III}, we optimize the decoherence rates for maximal excitation
energy transfer efficiency and explore how the efficiency depends on
the initial state of the FMO. The fluctuations in the local on-site
energy and in the site-site couplings affect the transfer
efficiency, this effect is studied in Sec. {\rm IV}. Finally, we
conclude our results in Sec. {\rm V}.

\section{model}
In the following, we will briefly introduce the model Hamiltonian
that we use to describe the FMO complex. For each site we take two
electronic states into account. Since we restrict ourself  to a
single electronic excitation, the so-called Frenkel exciton
Hamiltonian,
\begin{equation}
H=\sum_{j=1}^7 E_j|j\rangle\langle
j|+\sum_{i>j=1}^7J_{ij}(|i\rangle\langle j|+h.c),
\end{equation}
can describe the reversible dynamics of the electronic degrees of
freedom. Here $|j\rangle$ represents the state where only the $j$-th
site is excited and all other sites are in their electronic ground
state. $E_j$ is the on-site energy of site $j$, and $J_{ij}$ denotes
the excitonic coupling between sites $i$ and $j$. In the site basis,
we follow Ref.\cite{adolphs06} and employ the Hamiltonian matrix
elements (in units of $\mbox{cm}^{-1}$) in the remainder of this
paper,
%\begin{widetext}
{\tiny
\begin{equation}
        H \!=\!\! \left(\!\!\begin{array}{rrrrrrr}
         \mathbf{215}   & \!\mathbf{-104.1} & 5.1  & -4.3  &   4.7 & \mathbf{-15.1} &  -7.8 \\
        \!\mathbf{-104.1} &  \mathbf{220.0} &\mathbf{ 32.6} & 7.1   &   5.4 &   8.3 &   0.8 \\
           5.1 &  \mathbf{ 32.6 }&  0.0 & \mathbf{-46.8} &   1.0 &  -8.1 &   5.1 \\
          -4.3 &    7.1 &\!\mathbf{-46.8} & \mathbf{125.0} &\! \mathbf{-70.7} &\! -14.7 &
          \mathbf{ -61.5}\\
           4.7 &    5.4 &  1.0 & \!\mathbf{-70.7} & \mathbf{450.0} & \mathbf{ 89.7} &  -2.5 \\
         \mathbf{-15.1} &    8.3 & -8.1 & -14.7 &  \mathbf{89.7} & \mathbf{330.0} & \mathbf{ 32.7} \\
          -7.8 &    0.8 &  5.1 & \mathbf{-61.5} &  -2.5 &  \mathbf{32.7} & \mathbf{280.0}
          \end{array}\!\!
        \right).
        \label{ha}
\end{equation}
}
%\end{widetext}
Here the zero energy has been shifted by 12230 $\mbox{cm}^{-1}$ for
all sites, corresponding to a wavelength of $\sim 800 \mbox{nm}$. We
note that in units of  $\hbar=1$, we have 1 ps$^{-1}$=5.3 cm$^{-1}$.
Then by dividing $J_{ij}$ and $E_{j}$ by 5.3, all elements of the
Hamiltonian  are rescaled in units of ps$^{-1}$. We can find from
the Hamiltonian $H$ that in the Fenna-Matthew-Olson complex (FMO),
there are two dominating exciton energy transfer (EET) pathways:
$1\rightarrow 2\rightarrow 3$ and $6\rightarrow (5,7)\rightarrow 4
\rightarrow 3$. Although the nearest neighbor terms dominate the
site to site coupling, significant hopping matrix elements exist
between more distant sites. This indicates that coherent transport
itself may not explain why the excitation energy transfer is so
efficient.

We adopt the spin-boson model to describe the interactions between
excitations and surrounding protein environments,
\begin{equation}
H_{sb}=\sum_{i=1}^7\sum_j g_{ij}|i\rangle\langle
i|(a_j^{\dagger}+a_j),
\end{equation}
where $g_{ij}$ represents the coupling constant between site $i$ and
the mode $j$ of the environment. The Hamiltonian $H_{sb}$ describes
the modulation of on-site energy by the environment in the liner
case, where the on-site energy on the dimensionless environment
coordinate $q_j \sim (a_j^{\dagger}+a_j)$. Such a site-environment
coupling can model the site-vibration interaction, leading to
decoherence in the FMO. This decoherence can be described by the
master equation. To derive the master equation,  we adopt the
following assumptions. (1) The time evolution of the whole system
(the FMO + environment) is unitary, where the full Hamiltonian  is
assumed to be time-independent and consists of three parts, namely
the system Hamiltonian $H$, the bath Hamiltonian $H_b=\sum_j
\hbar\omega_j a^{\dagger}_j a_j$ and the interaction Hamiltonian
$H_{sb}$. The deviation of the master equation is equivalent to
finding the dynamics of the FMO by tracing out the degrees of
freedom of the environment. This is not always possible and we shall
assume that, (2) the system-environment interaction is sufficiently
weak, so that perturbation theory is applicable. Moreover, we assume
that, (3)  the whole system is in a product initial state, and (4)
the environment  has short memory in the sense that the correlation
time is very short. For details, we refer the readers to Ref.
\cite{breuer02,gardiner91}. With these assumptions, we can derive a
master equation to describe the dynamics of the FMO complex, which
is given by,
\begin{eqnarray}
        \frac{d\rho}{dt} = -i[H,\rho]
        + {\cal L}(\rho) + {\cal L}_{38}(\rho)\;, \label{masterE}
\end{eqnarray}
where the Liouvillian takes,
\begin{eqnarray}
\mathcal{L}(\rho)&=&\sum_{j=1}^7 \gamma_j \left[P_j \rho(t)
P_j-\frac{1}{2}P_j\rho(t)-\frac{1}{2}\rho(t)P_j\right],\nonumber\\
\mathcal{L}_{38}(\rho)&=& \Gamma \left[P_{83}\rho(t)
P_{38}-\frac{1}{2}P_3\rho(t)-\frac{1}{2}\rho(t)P_3\right],
\end{eqnarray}
with $P_j=|j\rangle\langle j|$, $j=1,2,3,...,7$, and
$P_{38}=|3\rangle\langle 8|=P_{83}^{\dagger}.$  $\mathcal{L}(\rho)$
describes exciton decoherence due to the site-environment couplings,
and  $\mathcal{L}_{38}(\rho)$ characterizes the excitation trapping
at site 3 due to interactions with the  reaction center, labeled as
site 8. The presence of this 8th BChl chromophore has been suggested
in each subunit of the FMO complex by recent crystallographic
data\cite{tronrud2009}. Furthermore, experimental data and
theoretical studies indicated that the 8th BChl is the closest  to
the baseplate and should be the point at which energy flows into the
FMO complex\cite{busch11,ritschel11}.

\begin{figure}
\includegraphics*[width=0.8\columnwidth,
height=0.5\columnwidth]{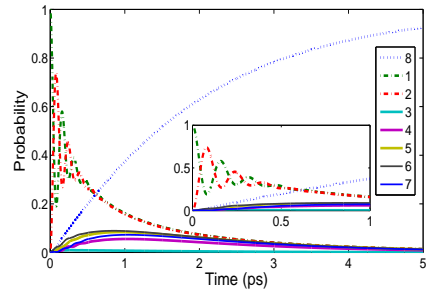} \caption{The excitation
transfer obtained from the master equation. The decoherence rates
are optimized for maximal transfer efficiency, which gives
$\gamma_1=1.2, \gamma_2=24, \gamma_3=0.9333, \gamma_4=6.5333,$ $
\gamma_5=53.1, \gamma_6=1.8, \gamma_7=27$ and $ \Gamma=44,$ with
   transfer efficiency  $p_8=0.9256.$  The exciton is assumed
initially on the  site 1.} \label{fig1}
\end{figure}
We shall use the population $p_8$ at time $T$ in the reaction center
given by $p_8(T)=\Tr(|8\rangle\langle 8|\rho(T))$ to quantify the
excitation transfer efficiency. Clearly, the Liouvillian
$L_{38}(\rho)$ plays an essential role in the excitation transfer.
Through this term the decoherence with rates $\gamma_j$
$(j=1,2,...,7)$ can enhance the excitation transfer as we show in
the next section.

\section{results}
It is believed that the completely coherent dynamics is not most
ideal for the excitation transfer, and the quantum coherence itself
cannot explain the very high excitation transfer efficiency. Indeed,
our numerical simulations show that $\Gamma$ can be optimized to
87.14 with $\gamma_j=0$ to obtain the    excitation transfer
efficiency $p_8=0.6781$.

\begin{figure}
\includegraphics*[width=0.8\columnwidth,
height=0.5\columnwidth]{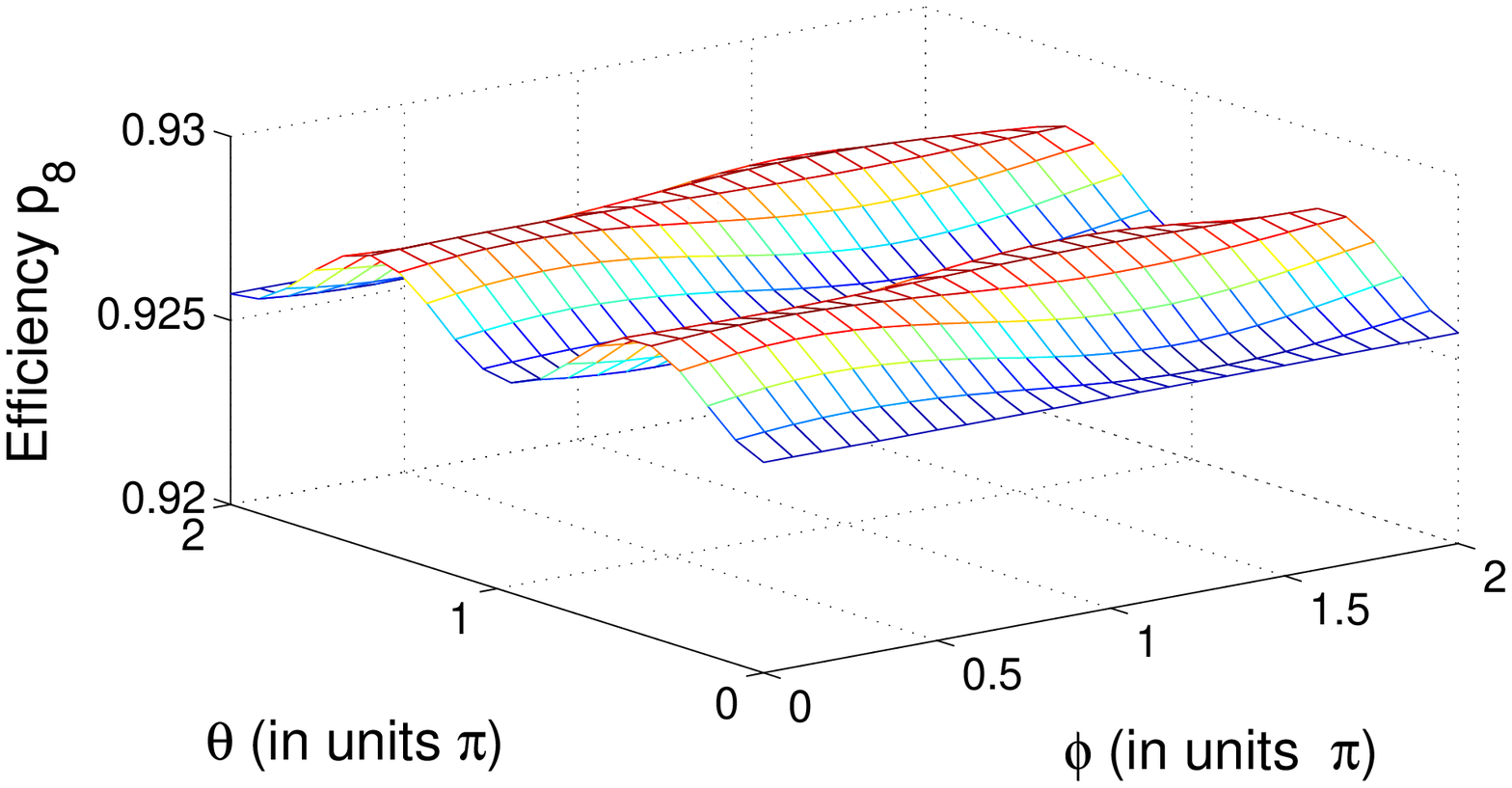}
\includegraphics*[width=0.8\columnwidth,
height=0.5\columnwidth]{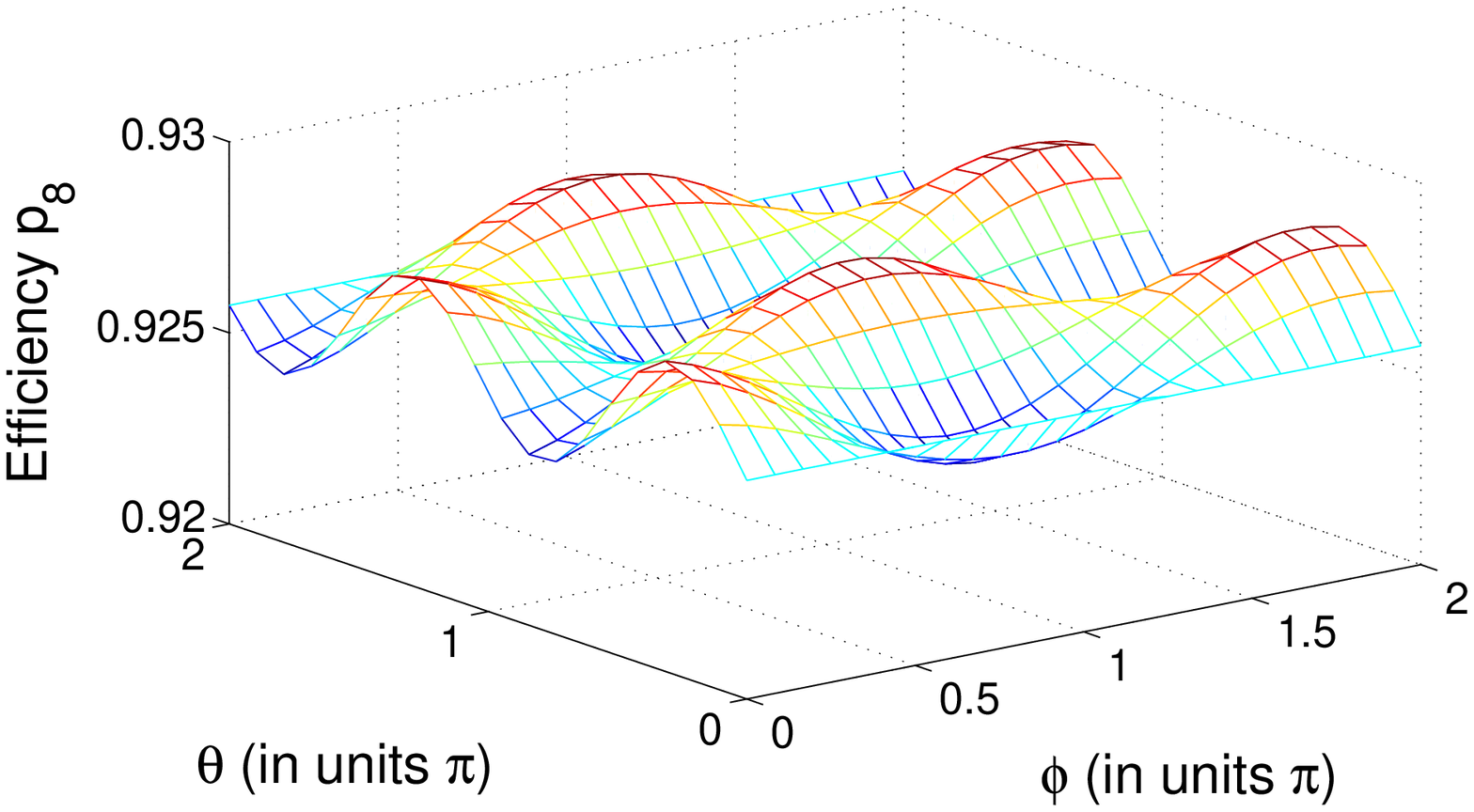}\caption{The dependence of the
transfer efficiency on the initial states. Top panel:  the initial
state site  is a superposition of $|1\rangle$  and $|6\rangle$.
Namely, $|\psi(t=0)\rangle=\cos\theta|1\rangle+\sin\theta\exp(i
\phi)|6\rangle.$ Lower panel: the initial state is
$|\psi(t=0)\rangle=\cos\theta|1\rangle+\sin\theta\exp(i
\phi)|2\rangle.$ The decoherence rates are optimized for maximal
transfer efficiency, i.e., they take the same values  as in
Fig.\ref{fig1}. 0.5;0.25,0.75} \label{fig2}
\end{figure}

We optimize the decoherence rates $\Gamma$ and $\gamma_j,
(j=1,2,...,7)$ for the maximal excitation energy transfer
efficiency, and find that $\Gamma=44$  and $\gamma_1=1.2,
\gamma_2=24, \gamma_3=0.9333, \gamma_4=6.5333,$ $ \gamma_5=53.1,
\gamma_6=1.8, \gamma_7=27$ yield a  transfer efficiency
$p_8=0.9256.$ With these decoherence rates, the population on each
site as a function of time is plotted in Fig. \ref{fig1}. Two
observations can be made from Fig. \ref{fig1}: (1) The population
shows oscillatory behaviors during the first 0.5ps and then they
change in a smooth way, (2) the excitation on site 1 and 2 dominates
the population, while the population at site 3 is almost zero. The
first observation can explain why the excitation transfer lasts for
longer time than the quantum coherence (quantum oscillation), and
the second observation suggests that the site 1 and site 2 play
important role in the excitation energy transfer. The excitation on
site 3 is trapped and transferred to the reaction center almost
immediately as to obtain a high transfer efficiency, as Fig.
\ref{fig1} shows.
\begin{figure}
\includegraphics*[width=0.8\columnwidth,
height=0.5\columnwidth]{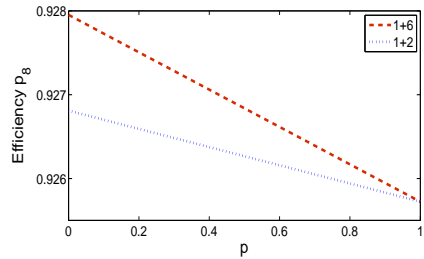} \caption{The transfer
efficiency versus initial states: the case of classical
superposition. The decoherence rates are the same as in Fig.
\ref{fig1}. The initial state is  $p|1\rangle\langle
1|+(1-p)|6\rangle\langle 6|$ for red-dashed line, and $p
|1\rangle\langle 1|+(1-p)|2\rangle\langle 2|$ for blue-dotted line.}
\label{fig3}
\end{figure}

Spatial and temporal relaxation of exciton  shows that the site 1
and 6 were populated initially with larger contribution
\cite{adolphs06}. Then it is interesting to study how the initial
states affect the excitation transfer efficiency. In the following,
we shall shed light on this issue. Two cases are considered. First,
we calculate the excitation transfer efficiency with exciton
initially in a superposition of site 1 and 6, and study the effect
of transfer efficiency on the initial states. Second, we extend this
study to initial states where sites 1 and 2 are initially excited.
These calculations are performed by numerically optimized the
decoherence rates for maximal transfer efficiency at time $T= 5 $ps,
selected results are shown in Fig. \ref{fig2}, \ref{fig3},
\ref{fig5}.

\begin{figure}
\includegraphics*[width=0.8\columnwidth,
height=0.5\columnwidth]{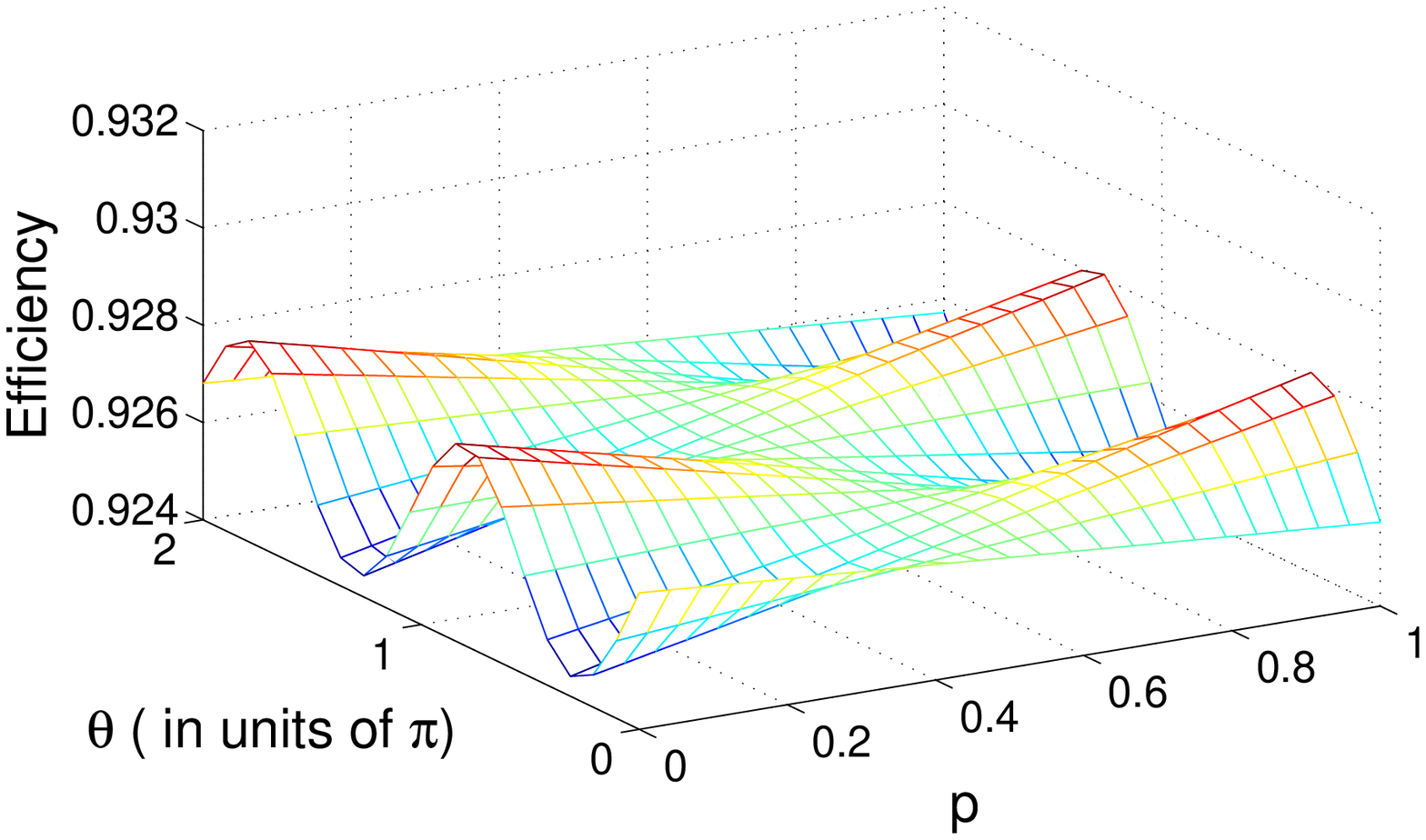}
\includegraphics*[width=0.8\columnwidth,
height=0.5\columnwidth]{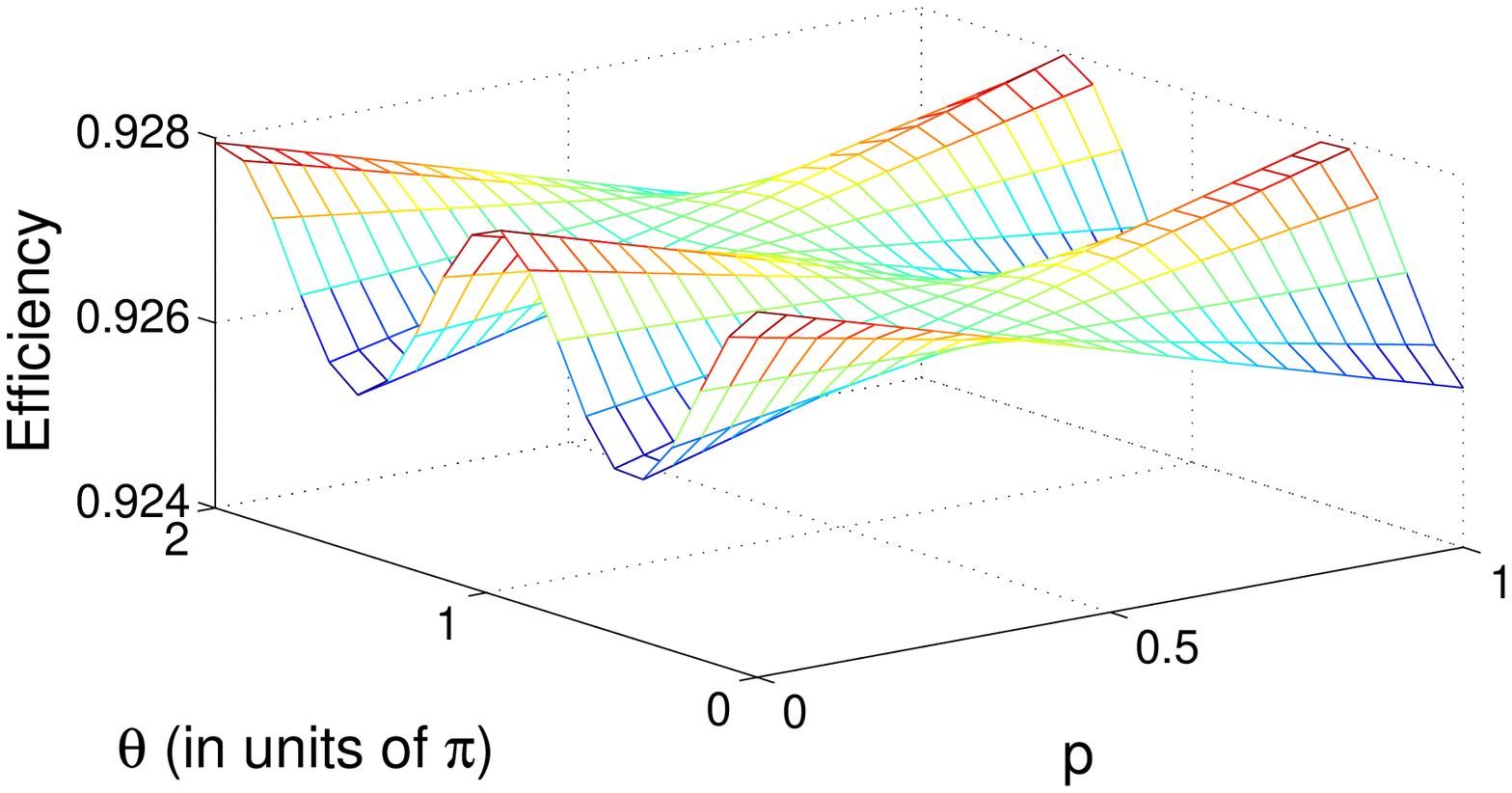} \caption{The dependence of
the transfer efficiency on the initial states. The decoherence rates
are optimized for maximal transfer efficiency. The upper  and lower
panel are for different initial states with $\phi=0$. Upper panel:
$\rho(t=0)=p|\psi_0\rangle\langle
\psi_0|+(1-p)|\psi_0^\bot\rangle\langle\psi_0^\bot|$ with
$|\psi_0\rangle=\cos\theta |1\rangle
+\sin\theta\exp(i\phi)|2\rangle$ and
$|\psi_0^\bot\rangle=\cos\theta\exp(i\phi)|2\rangle-\sin\theta|1\rangle$.
Lower panel: $|2\rangle \longleftrightarrow |6\rangle.$}
\label{fig4}
\end{figure}
In Fig. \ref{fig2} we present the transfer efficiency as a function
of $\theta$ and $\phi$, which characterize the pure initial states
of the FMO complex through
$|\psi(t=0)\rangle=\cos\theta|1\rangle+\sin\theta\exp(i
\phi)|6\rangle$ (upper panel) and
$|\psi(t=0)\rangle=\cos\theta|1\rangle+\sin\theta\exp(i
\phi)|2\rangle$ (lower panel). We find that a properly coherent
superposition of sites 1 and 6 (or 2) can enhance the exciton
transfer. For exciton initially in a superposition of 1 and 6, the
transfer efficiency is more sensitive to the population ratio
(characterized by $\theta$) but not to the relative phase $\phi$.
The transfer efficiency arrives at its maximum with $\theta=\frac
\pi 2$ and $\phi=0$ (it is not unique), suggesting that exciton on
site 6 is more favorable for the energy transfer. This finding is
changed when the relative phase $\phi$ is not zero, for example,
when $\phi=\pi$, the maximal transfer efficiency moves toward larger
$\theta$.  For exciton initially excited on site 1 and 2, both the
population ratio and the relative phase affect the energy transfer,
the transfer efficiency reaches its maximum with $\theta=\frac \pi
4$ and $\phi=0$ and its minimum with $\theta=\frac {3\pi} {4}$ and
$\phi=0$.

Fig. \ref{fig3} shows the dependence of the transfer efficiency on
the  mixing rate $p$, where the initial state is a classical mixing
of site 1 and 6 (or 2). Obviously, the population mixing favors the
transfer efficiency. In particular, the classical mixing of site 1
and site 6 (or site 2) always increase the transfer efficiency, this
is different from the case where the mixing is performed  in a
different way, see Fig. \ref{fig4}. We observe from Fig. \ref{fig4}
that the transfer efficiency attains  a minimal value at
$\theta=\frac 3 4 \pi$ and $p=1$ (upper panel) which is consistent
with the observation emerged from Fig. \ref{fig2}. As $p$ decreases,
the transfer efficiency increases linearly. Similar features can be
found from the lower panel of Fig. \ref{fig4}. These results suggest
that the population on site 6 favors the energy transfer: if the
mixing of site 6 and site 1 is classical, the transfer efficiency
increases linearly with the population on site 6. Otherwise, it
depends on the mixing angle $\theta$ and the relative phase $\phi$.

We would like to note that the coherent superposition of  exciton on
the sites 1 and  6 (or 2) can  help the energy transfer even when
the  decoherence is absent. For example,  our numerical
optimizations show that the transfer efficiency can be increased to
98\% starting with an initial state
$\cos(1.2632\pi)|1\rangle+\sin(1.2632\pi)|2\rangle$, however, this
is not the case for classical mixing (or incoherent superposition)
of the sites 1 and 2 (or site 6) as initial states.  The transfer
efficiency is always smaller than (or equal to) 67.81\% (the
efficiency in the case with exciton 100\% on the site 1 in the
absence of decoherence) with  initial states in the form
$p|1\rangle\langle 1|+(1-p)|2\rangle\langle 2|$. This suggests that
the transfer enhancement by coherent superposition of exciton on
site 1 and site 6 (or site 2) is a common feature for the energy
transfer in the FMO complex regardless of the presence of
decoherence. But  the enhancement of transfer efficiency by
classical mixing works only in the presence of decoherence. This can
be understood as the coherent cancelation or coherent enhancement in
the dynamics of exciton transport.
\begin{figure}
\includegraphics*[width=0.8\columnwidth,
height=0.5\columnwidth]{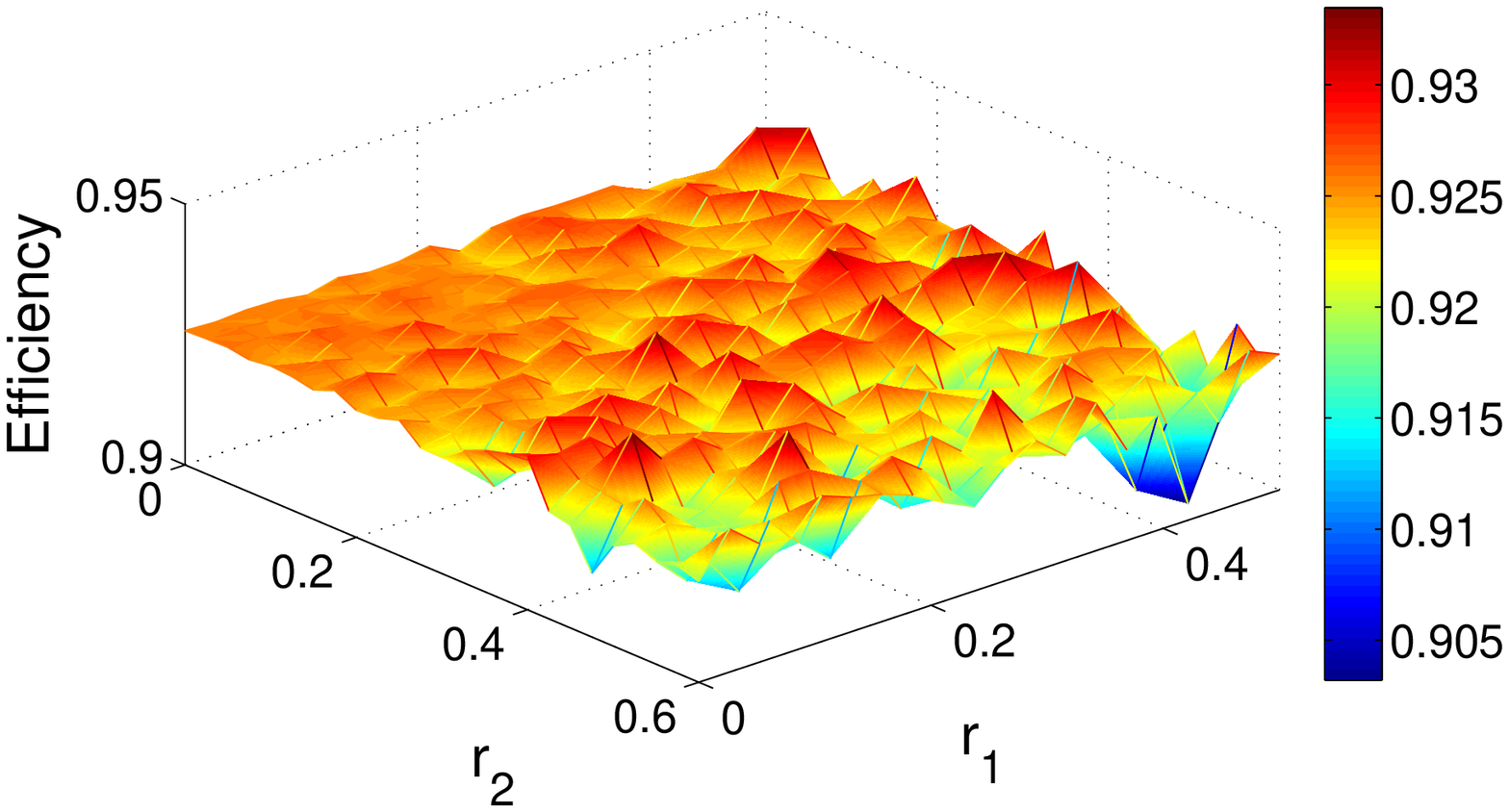}
\includegraphics*[width=0.8\columnwidth,
height=0.5\columnwidth]{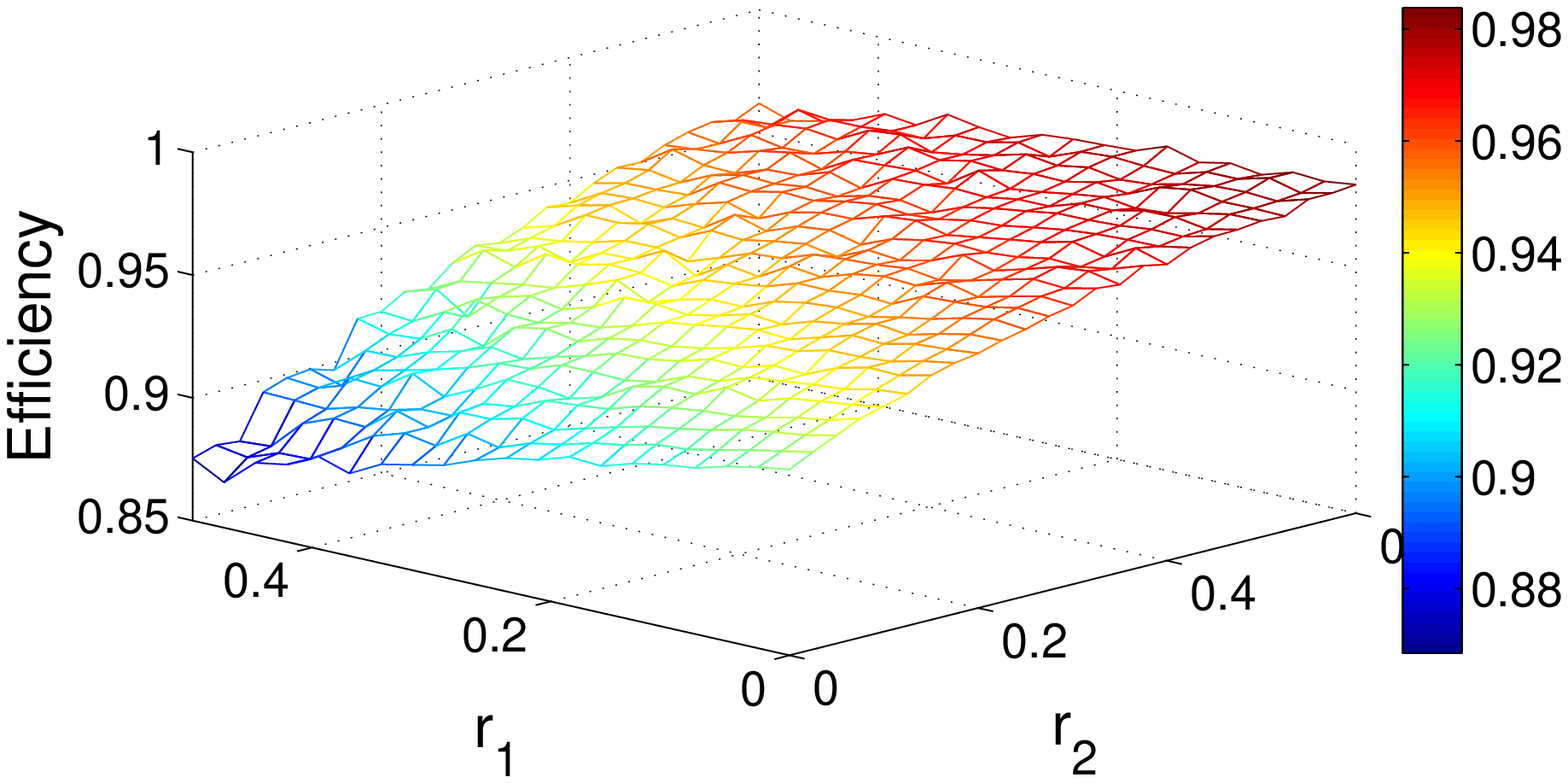}\caption{Effect of
fluctuations in the Hamiltonian on the transfer efficiency. The
upper panel is for  the fluctuations with zero mean, while the lower
panel is for fluctuations with positive mean. The resulting fidelity
is an averaged result  over 20 independent runs. } \label{fig5}
\end{figure}

\begin{figure}
\includegraphics*[width=1.1\columnwidth,
height=0.5\columnwidth]{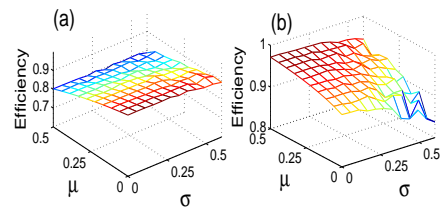} \caption{Effect of
fluctuations in the Hamiltonian on the transfer efficiency.  The
fluctuations are Gaussian with mean $\mu$  and the standard
deviation $\sigma$. (a) The fluctuations happen only in the on-site
energies, (b) the fluctuations in both the on-site energies and
couplings. The resulting fidelity is an averaged result over 100
independent runs.} \label{fig9}
\end{figure}
It has been shown that the local decoherence can enhance the
transfer efficiency \cite{caruso09} in the FMO, this can be
understood as  the fluctuation-induced-broadening   of energy
levels, which bridges the on-site energy gap and the coupling
between them. This gives rise to the following question:  how the
fluctuations in the site  energies and couplings  affect the
transfer efficiency. In the following, we will study this issue and
show that the exciton transfer in the FMO complex remains
essentially unaffected in the presence of random variations in site
 energies and inter-site couplings.  This strongly suggests that the
experimental results recorded for samples at low temperature would
also be observable at higher temperatures.

We take the decoherence rates that maximize the transfer efficiency
for the numerical simulation. Two types of fluctuations in the site
energies and inter-site couplings  are considered. In the first one,
it has zero mean, while the another type of fluctuations has nonzero
positive mean. For the fluctuations with zero mean, the Hamiltonian
in Eq. (\ref{ha}) takes the following changes, $H_{jj}\rightarrow
H_{jj}(1+r_1\cdot(\mbox {rand}(1)-0.5))$ and $H_{i\neq j}\rightarrow
H_{i\neq j}(1+r_2\cdot(\mbox {rand}(1)-0.5)).$ For the fluctuations
with nonzero positive mean, the Hamiltonian takes the following
changes, $H_{jj}\rightarrow H_{jj}(1+r_1\cdot \mbox {rand}(1))$  and
$H_{i\neq j}\rightarrow H_{i\neq j}(1+r_2\cdot\mbox {rand}(1)),$
where $\mbox{rand(1)}$ denotes a random number between 0 and 1. So a
100\% static disorder   may appear in the on-site energies and
inter-site couplings.

With these arrangements, we numerically calculated the transfer
efficiency and present the results in Fig. \ref{fig4}. Each transfer
efficiency is a result averaged over 20 fluctuations. Two
observations are obvious. (1) As the fluctuations with zero-mean in
the site energies and couplings increase, the transfer efficiency
fluctuates greatly. (2) For fluctuations with nonzero  positive
mean, the efficiency increases with $r_2$ but decrease with $r_1$.
This can be interpreted as follows. The energy gap between
neighboring sites blocks the energy transfer, whereas the inter-site
couplings  that represent the overlap of different sites favor the
transport. As a result of competition, the efficiency increases with
$r_1$ but decreases with $r_2$. For the fluctuation with zero mean,
the results are subtle, it depends seriously on each run, and the
averaged transfer efficiency remains almost  unchanged.

Ref.\cite{adolphs06} suggested that the fluctuations in the on-site
energies are Gaussian. To examine the effect of Gaussian
fluctuations, we introduce a Gaussian function,
$y(x|\sigma,\mu)=\frac{1}{\sigma\sqrt{2\pi}}e^{-\frac{(x-\mu)^2}{2\sigma^2}},$
where $\mu$ is the mean, while $\sigma$ denotes the standard
deviation. Furthermore we assume that the fluctuations enter into
only the on-site energies, namely,  $H_{jj}$ is replaced by
$H_{jj}(1+y(x|\sigma,\mu))$, $j=1,2,...,7$. With these notations, we
plot the effect of fluctuations on the transfer efficiency in Fig.
\ref{fig9}(a). We find that the efficiency is almost independent of
the variation, but it decreases as the mean increases. This again
can be understood as the blockage of energy transfer by the energy
gap between the neighboring sites. When these Gaussian fluctuations
occur in both the on-site energies and the couplings, we find from
Fig.\ref{fig9}(b) that the transfer efficiency increases as the mean
increases and the variation decreases. This is a result of
competition between the energy gaps  and  couplings of the
neighboring sites.

Some one may wonder about the decoherence rates that yield the high
energy transfer efficiency -- why the optimal decoherence rate is
different for each sites? why some decoherence rates are larger than
the energy gaps between the sites. As a phenomenal model presented
here, we can not answer these questions, this stimulate further
works on this topic.

\section{Conclusion}
In this paper, we  study the excitation transfer efficiency  in
light-harvesting complexes, in particular, the Fenna-Matthew-Olson
(FMO) complex. This is a problem of long interest, and timely due to
the recent activity sparked by suggestions of quantum effects in
energy transport. It is well known that photosynthetic organisms
operate at extremely low light levels, and the exciton initially
excited in the FMO complex may occupy the site 1 and 6 (or site 2)
simultaneously. This stimulate us to study the effects of initial
state on the transfer efficiency. Based on the Frenkel exciton
Hamiltonian and the master equation, we have optimized the
decoherence rates for maximal energy transfer efficiency, which can
reach about 93\% at time 5ps.  By considering different mixing of
exciton on site 1 and site 6 (site 2) as the initial states, we have
examined the effect of initial states on the energy transfer
efficiency. The results suggest that a classically mixing of site 1
and site 6 (or site 2)  always enhance the energy transfer
efficiency, whereas a coherent superposition of site 1 and site 6
(or site 2) increases the efficiency only for special initial
states. This can be understood as coherent cancelation or coherent
enhancement in quantum physics.

\ \ \\
This work is supported by the NSF of China under Grants No 61078011,
No 10935010, and No 11175032 as well as the National Research
Foundation and Ministry of Education, Singapore under academic
research grant No. WBS: R-710-000-008-271.

\end{document}